\documentclass[12pt,preprint]{aastex}
\begin {document}

\title{Spectroscopic and Spectropolarimetric Observations of V838 Mon}

\author{John P. Wisniewski\altaffilmark{1}, Nancy D. Morrison\altaffilmark{1}, Karen S. Bjorkman\altaffilmark{1}, Anatoly S. Miroshnichenko\altaffilmark{1}, Amanda C. Gault\altaffilmark{1}, Jennifer L. Hoffman\altaffilmark{2,3}, Marilyn R. Meade\altaffilmark{4}, \& Jason M. Nett\altaffilmark{4}}

\altaffiltext{1}{Ritter Observatory, MS \#113, Department of Physics and Astronomy, University of Toledo, Toledo, OH 43606-3390 USA, jwisnie@physics.utoledo.edu, nmorris2@uoft02.utoledo.edu, karen@physics.utoledo.edu, anatoly@physics.utoledo.edu,
agault@utphya.panet.utoledo.edu}

\altaffiltext{2}{Department of Astronomy, University of Wisconsin-Madison, 475 N. Charter St., Madison, WI 53706}

\altaffiltext{3}{Department of Physics and Astronomy MS-108, Rice University,
6100 Main Street, Houston, TX 77005, jhoffman@rice.edu}

\altaffiltext{4}{Space Astronomy Lab, University of Wisconsin-Madison,
1150 University Avenue, Madison, WI 53706, meade@sal.wisc.edu,
jnett@sal.wisc.edu}

\begin{abstract}

The spectroscopic and spectropolarimetric variability of the peculiar
variable V838 Monocerotis during the brighter phases of its multiple
 outbursts in 2002 is presented.
Significant line profile variability of H$\alpha$ and Si II
6347.10\AA\ \& 6371.36\AA\ occurred in spectra obtained between 2002 February 5 
and 2002 March 14, and a unique secondary absorption component was observed
near the end of
this time period.  Our observations also suggest that multiple shifts 
in ionization states occurred during the outbursts.
Spectropolarimetric observations reveal that V838 Mon exhibited 
both intrinsic and interstellar polarization components during the initial
stages of the second
 outburst, indicating the presence of an asymmetric geometry;
however, the intrinsic component had significantly declined by February 14.
  We determine
the interstellar polarization to be $P_{max} = 2.746 \pm 0.011 \%$,
$\lambda_{max} = 5790 \pm 37\AA$, $PA = 153.43 \pm 0.12 ^{\circ} $, and we find
the integrated intrinsic V band polarization on February 5 to be  $P = 0.983
 \pm 0.012 \%$ at a position angle of $127.0 \pm 0.5^{\circ}$.
The implications of these observations for the nature of V838 Monocerotis,
its distance, and its ejecta are discussed.

\end{abstract}


\section{Introduction}

\citet{bro02} reported the discovery of a possible nova, later
to be designated V838 Monocerotis, on 2002 January 6.6.  Prior to
outburst, V838 Mon was a hot blue star,
whose B band brightness was stable at 15.85 $\pm$ 0.4
 from 1949-1994 \citep{gor02}.
\citet{mu102} noted that V838 Mon was not detected in prior
H$\alpha$ emission-line surveys.  A spectrum obtained on January 26  
showed numerous neutral metal and s-process lines, and resembled
 that of a heavily reddened K-type giant \citep{zwi02}.

V838 Monocerotis underwent a second major photometric outburst in 
early February 2002, changing from V=10.708 on February 1.86 to
V=8.024 on February 2.91 \citep{kim02}, and reaching a maximum brightness
of 6.66 in V on February 6 \citep{gor02}.  Spectra obtained during and
immediately following this outburst \citep{iij02,mor02} revealed the
emergence of various ionized metal lines.  \citet{kae02} estimated a
blackbody continuum temperature of 4500 K to be present on 
February 9, and \citet{hen02} found that a light echo had developed around 
V838 Mon on February 17.  IRAS source 07015-0346 has been associated with
the location of V838 Mon \citep{kat02}, leading to the suggestion that this
IR emission is from the dust causing the light echoes \citep{kim2b}.

A third, less intense outburst occurred in early March 2002 \citep{kim2b, mu102}.
By April 16, V838 Mon's spectrum had evolved such that it resembled 
an M5 giant \citep{rau02} , with strong TiO molecular bands and a
temperature of $\sim$3000 K.  \citet{ban02} detected TiI emission lines from
near infrared spectroscopy, peaking in strength on May 2, and argued that this
emission arose from circumstellar ejecta.  They used the strengths of these TiI
lines to estimate the mass of V838 Mon's ejected envelope to be 10$^{-7}$ to
10$^{-5}$ M$_{\sun}$.

By October 2.17, spectroscopic observations suggested that V838 Mon had evolved
into a ``later than M10 III'' type star \citep{des02}.  \citet{des02} also
 detected a weak blue continuum, suggesting
the presence of a binary companion.  Followup spectroscopy 
 \citep{wag02,mu302} confirmed this detection, and \citet{mu302}
 suggested that
the companion was a B3 V type star.  The unique, complex evolution of V838 
Monocerotis has led \citet{mu102} to suggest that this object represents a new
class of objects, ``stars erupting into cool supergiants (SECS)''.

In this paper, we report the spectroscopic and spectropolarimetric
properties of V838 Monocerotis following its second and third
photometric outbursts.  In section 2, we outline
our observational data.  We detail the equivalent width and line profile
variability of selected spectral lines in section 3.1.  Our
spectropolarimetric data, most notably the detection of an intrinsic
polarization component, are discussed in section 3.2.  We address the
distance to V838 Mon in section 3.3.  Finally, in
section 4, we discuss the implications of these observations for
future studies of this unique object.

\section{Observations}

We obtained spectroscopic observations of V838 Monocerotis with the
Ritter Observatory 1m reflector, using a fiber-fed echelle spectrograph.
The fiber used for these observations has a diameter of 200 $\mu$m,
which corresponds to roughly 5$^{''}$ on the sky.
Nine non-adjacent orders of width 70 \AA\ were observed in the range
 5285 -6595 \AA.  Data were recorded on a
 1200 x 800 Wright Instruments Ltd. CCD, with 22.5 x 22.5 $\mu$m pixels.  
With $R \equiv \lambda /\Delta \lambda \simeq 26,000,$ the spectral 
resolution element, $\Delta \lambda$, is about 4.2 pixels owing to a 
widened entrance slit.  
  Observations were reduced with IRAF\footnote{IRAF
is distributed by the National Optical Astronomy Observatories, which are
operated by the Association of Universities for Research in Astronomy, Inc.,
under contract with the National Science Foundation.}
 using standard techniques.  Further details about the reduction of Ritter data
can be found in \citet{mor97}.  Unless otherwise noted, all data 
were shifted to the
heliocentric rest frame and continuum normalized.  

We obtained spectropolarimetric observations of V838 Mon with the University of
Wisconsin's HPOL spectropolarimeter, which is the dedicated instrument
 on the 0.9m Pine
Bluff Observatory (PBO) telescope.  These data were recorded with a 400
x 1200 pixel CCD camera, covering the wavelength range of 3200
-10500\AA, with a spectral resolution of 7\AA\ below 6000\AA\ and
10\AA\ above this point \citep{nor96}.  Observations were made with dual 6
x 12 arc-second apertures, with the 6 arc-second slit aligned E-W and
the 12 arc-second decker aligned N-S on the sky.  The two apertures allow 
simultaneous star and sky data to be recorded, providing a reliable means for
subtraction of background sky polarization and hence allowing accurate
 observations to be made even in non-photometric skies.  We processed these data
using REDUCE, a spectropolarimetric software package developed by the
University of Wisconsin-Madison \citep{wol96}.  Further details about
HPOL and REDUCE may be found in \citet{noo90} and \citet{har00} .
Instrumental polarization is monitored on a weekly to monthly basis at PBO
 via observations
of polarized and unpolarized standard stars, and over its 13 year existence,
 HPOL
has proven to be a very stable instrument.  We have corrected our data for 
instrumental effects to an absolute accuracy of $0.025\%$ and 1$^{\circ}$
 in the V band (Nordsieck, private communication).  
 HPOL spectroscopic data are not calibrated to an absolute
 flux level due to the non-photometric skies routinely present \citep{har00} .

Table 1 provides a log of the observations from both observatories.

\section{Results}

\subsection{Spectroscopic Variability}

We now discuss the spectral evolutionary history of V838
 Monocerotis from February 5 to March 14.
Our observations of V838 Mon from February 5 to February 9, during
 the onset of the second photometric outburst, indicate an
overall shift toward a higher ionization state \citep{iij02,mor02} as
compared with initial observations \citep{zwi02}.   

H$\alpha$ shows a strong P-Cygni profile, with electron scattering wings
extending at least $\pm$ 1100 km s$^{-1}$ from February 5 to February 8 and 
about 850 km s$^{-1}$ on February 9, and an average heliocentric
blue edge radial velocity of -300 km s$^{-1}$ (see Figure 1).  This radial
velocity is slightly lower than the terminal velocity of -500 
km s$^{-1}$ observed in late January in CaII, BaII, NaI, and LiI lines
 \citep{mu202}.  \citet{gor02} report that a spectrum on February 5 shows
H$\alpha$ with FWZI = 3100 km s$^{-1}$ and an absorption component at -300
 km s$^{-1}$, which is inconsistent with our findings.  The extent of
 the electron scattering wings strongly depends on accurate continuum
placement.  We are confident that, within the limits of the SNR of our
data, we see a 5 \AA\ ``flat'' continuum region at each end of the spectral
interval containing H$\alpha$, hence we are accurately determining the continuum
level.  The total equivalent width peaked on February 6 (see
 Table 3), and then began a 
steady decrease.   Equivalent width errors were calculated using
 $\sigma ^{2} = N (h_{\lambda} / SNR)^{2} (f_{*} /
f_{c})$, where N is the number of pixels across a line,
$h_{\lambda}$ is the dispersion in \AA\ pixel$^{-1}$, $f_{*}$ is the flux
in the line, $f_{c}$ is the flux at the continuum, and SNR is the signal
to noise ratio \citep{cha83}. 

In early February, all the strong lines exhibited significant line profile
variability.  In H$\alpha$ (Figure 1), the emission peak migrated to longer
wavelengths with time.  The high velocity component of Si II 6347.1\AA\ and
 6371.4\AA\ (Figure 2) weakens with time, and the instrinsic component of
 Na I 5889.95\AA\ and 5895.9\AA\ (Figure 3) also shows variability.  Note
that since the interstellar Na I components appear to be saturated, they could
not fit with gaussians and subtracted to reveal the pure intrinsic component.
The low resolution HPOL spectrum (Figure 4) on February 8 clearly depicts
the P-Cygni profiles of FeII 4923.9\AA, 5018.4\AA, 5169.0\AA\ and the
 CaII infrared triplet 
8498.0\AA, 8542.1\AA, 8662.1\AA.   Hydrogen Paschen absorption lines
at 8438.0\AA, 8467.3\AA, 8598.4\AA, 8750.5\AA, 8862.8\AA,
9014.9\AA,
9229.0\AA\ are observed, as well as HI 10049.4\AA, which has a clear
P-Cygni profile.

By February 14, the P-Cygni profile of H$\alpha$ had
weakened considerably(Table 3, Figure 1) and its absorption
and emission components were approaching equality in strength.  The strong
electron scattering wings previously observed had disappeared by our 
February 14 observation.  \citet{gor02} noted that the H$\alpha$ electron
scattering wings had disappeared in their spectrum taken on February 16. 
These results are consistent with a decreasing excitation level in the
circumstellar envelope.  A low
resolution red HPOL spectrum, obtained on February 13 (Figure 6), reveals
two other qualitative changes: the emission components of both the P-Cygni
CaII infrared triplet lines and HI 10049.4\AA\ line significantly decreased
in strength. 

By the end of the third photometric outburst, a significant shift in 
V838 Mon's spectral characteristics had occurred.  Specifically, our
spectra on March 11 showed that a second, high velocity absorption component
had developed in a few lines.  H$\alpha$ (Figure 1) clearly shows this
component, centered at a radial velocity of -200 km s$^{-1}$ with a blue edge
radial velocity of -280 km s$^{-1}$.  Figure 2 shows that this feature is also
present in the SiII 6347.1\AA\ line, centered at -200 km s$^{-1}$ with a blue
edge radial velocity of -260 km s$^{-1}$, and in the SiII 6371.4\AA\ line,
centered at -140 km s$^{-1}$ with a blue edge radial velocity of -190
 km s$^{-1}$.  Based
upon the radial velocities of these dual absorption features, we identify
the enormous P-Cygni profile around 6394\AA, seen in Figure 2, as FeI
6393.6\AA.  A strong P-Cygni profiled line in the vicinity of
6190\AA, which we
attribute to FeI 6191.6\AA, also emerged on March 11 (Figure 5).  
Figure 1 also reveals new spectroscopic features at 6544.9\AA, 6577.2\AA,
 and 6582.5 \AA, which we attribute to MgII 6545.9\AA\ and CII 6578.1\AA\
 and 6582.9\AA.  The apparent emergence of both higher excitation lines
 (CII and MgII) simultaneously with lower excitation lines (FeI) illustrates
 the complexity of V838 Mon's outburst.  In
fact, nearly all 9 orders of our spectra show
evidence for the emergence of new spectral features 
on March 11 and March 14.  Due to low signal to noise ratios, 
 as well as uncertainties in line blending and profile
shapes, we are unable to identify all lines definitively. 
 Since many of
these lines are consistent with the rest wavelengths of FeI, NeI, NiI,
TiII, MgII, and FeII, and since as noted above we have positively identified
two lines of FeI emerging on March 11, these results
indicate that V838 Mon began to experience a shift to a lower ionization state.
The evidence for spectral evolution that we observed will need to be combined
with that of other authors to portray a comprehensive picture of V838 Mon's
outbursts.

\subsection{Spectropolarimetric Variability}

Figures 5-6 illustrate the wavelength dependent polarization of V838
Mon on February 8 and February 13 respectively.  The differences between
these two observations are readily apparent.  The integrated Johnson R band
polarization of the February 8 data is $P = 3.226 \pm 0.004\%$ at a
position angle of $149.0 \pm 0.1^{\circ}$, while the R band polarization
 of the later
observation is $P = 2.667 \pm 0.004\%$ at a position angle of
$153.4 \pm 0.1^{\circ}$.
This change strongly suggests the presence of an intrinsic polarization
component.  Furthermore, the February 8 data are characterized by
strongly depolarized emission lines, while the February 13 polarimetric
data show no line features.  Polarimetric studies of Be stars
\citep{har68, coy76} have found that in contrast to continuum photons,
line emission, which predominantly originates in Be circumstellar disks,
has a low probability of being scattered.  With a few exceptions
\citep{mcl79,qui97}, emission lines should show little to no intrinsic
polarization.  Thus an intrinsically polarized emission line star should
exhibit depolarized emission lines, e.g. a superposition of polarized
continuum flux and unpolarized line flux.
 If one employs a similar argument with the ejecta of
V838 Mon, the strongly depolarized emission lines of February 8 may be
used to infer the interstellar polarization component (ISP).  Similarly, 
the absence of depolarization effects in the February 13 data
suggests that this polarization minima may be primarily attributed to
interstellar polarization.  
 As previously noted, the electron
scattering wings of H$\alpha$ disappeared by February 14.  Since one
expects electron scattering in the ejecta of V838 Mon to be the primary
source of any intrinsic polarization, the disappearance of the electron
scattering wings is consistent with the hypothesis that the
polarization signal observed on February 13 is primarily interstellar in nature.

In order to parametrize the wavelength dependence of the interstellar
 polarization in
the February 13 data, we fitted the empirical Serkowski law \citep{ser75},
as modified by \citet{wil82} to these data.  The resulting ISP
parameters are: $P_{max} = 2.746 \pm 0.011 \%$, $\lambda_{max}
= 5790 \pm 37\AA$, $PA = 153.43 \pm 0.12^{\circ}$, $\delta PA = 0$, and $K =
0.971$.  This fit is overlaid in Figures 5 -6.  This
Serkowski fit provides a near perfect fit to the 
February 13 observation; furthermore, it nicely fits the depolarized
emission lines in the February 8 observation.  We qualitatively crosscheck
this claim by using the polarization and extinction relationship,
formulated by \citet{ser75}, $3 E_{B-V} \le P_{max} \le 9 E_{B-V}$.
\citet{mu202} established a lower limit for the interstellar reddening
of E$_{B-V} \sim 0.25$ and suggested that the finding of \citet{zwi02},
 E$_{B-V}$ = 0.80 $\pm$ 0.05, represents an upper limit.  Following
the arguments of \citet{mu202}, we adopt the midpoint of these values,
E$_{B-V}$ = 0.50, which bounds the interstellar polarization along the line
of sight to V838 Mon by 
$1.5\% \le P_{max} \le 4.5\%$ , and thus qualitatively agrees with our
ISP determination.  \citet{mu202} reported
preliminary polarimetry results in which they suggested the ISP is
characterized by $P_{max} = 2.6\%$ at 5500 \AA\ at a position angle of
$150 \pm 2^{\circ}$.  We are thus confident that our parametrization
accurately describes the interstellar polarization component.

We used these Serkowski parameters to remove the ISP
component from the February 8 data, as seen in Figure 4,
leaving only the intrinsic component.  We find the integrated V
band intrinsic polarization to be $P = 0.983 \pm 0.012
\%$ at a position angle of $127.0 \pm 0.5^{\circ}$.  
It is interesting that the intrinsic polarization is clearly not wavelength
independent, which one would expect in the case of pure electron
scattering.  Rather, the polarization gradually increases at wavelengths
shortward and longward of $\sim$ 8000\AA, which suggests the presence
of  an absorptive
opacity source in V838 Mon's ejecta.  A possible Paschen jump, albeit only at
a one-sigma detection level, is visible in the raw and intrinsic
polarization in Figure 4.  Combined with the spectroscopic
observations of strong H$\alpha$ electron scattering wings on 
February 8, this might suggest a not unlikely speculation that hydrogen is
 the opacity source \citep{woo96,woo97}.

\subsection{Distance Estimations}

The distance to V838 Mon has yet to be
agreed upon.  \citet{mu102, mu202} followed the propagation of V838 Mon's
light echo, assuming a spherical distribution of scattering material, to
derive a distance of 790 $\pm$ 30 pc.  \citet{kim2b} used the same technique
on a different data set to estimate a distance of 640 to 680 pc.  \citet{bon02}
estimated a distance of 2.5 kpc from HST light echo images; however, 
it has been suggested that the geometry assumed by these authors is
unrealistic \citep{mu202, kim2b}.
  More recently, the reported detection of a hot binary
companion \citep{des02, wag02, mu302} has led \citet{mu302} to suggest a
distance of 10-11 kpc, based upon spectrophotometric parallax.
  We add to the above discussion by considering the
distance implied by our spectroscopic and polarimetric observations.

Based upon the assumption that cataclysmic variables contain no
intrinsic polarization, \citet{bar96} suggested a rough relationship
between polarization and distance.  When applied to sources near the
Galactic Plane, for distances $\le$ 1 kpc, this relation is given by
$P/d = 3.6\%$ kpc$^{-1}$.  Given our estimate of $P_{max}$ of $2.746\%$, this
would suggest a distance to V838 Mon of 763 pc. 

V838 Mon's strong, double interstellar Na I D lines provide a different
constraint on the distance.  At galactic longitude 217.8$^\circ$, radial
velocities of objects outside the solar circle are positive and increase
monotonically with increasing distance from the sun.  Thus, the radial
velocity of the longer wavelength component provides
a lower limit on the distance to V838 Mon.  The
 radial velocities of the two components of the
D lines were measured in the spectra of February 5, 6, 8, and 9. For D1
and D2, the means and standard deviations were, respectively, $21.9 \pm
0.6$, $22.1 \pm 0.8$, $47.9 \pm 0.8$, and $47.5 \pm 2.8$ km s$^{-1}$, relative
to the LSR.   
Note that our data are accurate to 2 km s$^{-1}$, as compared to the IAU
velocity standard $\beta$ Gem, which is constant to better than 0.1
km s$^{-1}$ \citep{lar93}.

To read off the distance of the 48 km s$^{-1}$, further cloud, we used the
velocity contour map by \citet{bra93}, which does not assume
the velocity field of Galactic rotation to be axisymmetric. The galactic
longitude of V838 Mon coincides with an interesting feature in this map,
an ``island'' of high velocities of about 50 km s$^{-1}$ located about 2500 pc
from the Sun. We estimate that distances consistent with this velocity
map, for a radial velocity of +48 km s$^{-1}$, lie in the range $2500 \pm
300$ pc. This estimate constitutes our lower limit on the distance to
V838 Mon. Velocities as large as 50 km s$^{-1}$ are not reached again in this
direction at heliocentric distances less than 8 kpc.  Since this lower limit
is greater than 1 kpc, the distance estimation technique used with our
polarimetric data is no longer applicable.

\section{Discussion}

Our spectroscopic data offer
both qualitative and quantitative insight into the initial stages of the
2002 outburst.  Future modeling efforts can be constrained by the
equivalent width variability of the lines presented.
Furthermore, the complex line profile variability and evolution of
various species and ionization stages of lines presented in this paper
should also provide constraints on future attempts to explain this outburst.

In spite of our sparse polarimetric data set, 
these observations clearly demonstrate that the
ejecta of V838 Monocerotis deviated significantly from a spherical
geometry.  We note the similarity between our observations and those of
\citet{bjo94}, who found Nova Cygni 1992 to have an intrinsic
polarization signal during the initial stages of outburst.  These
authors suggest the intrinsic polarization during this initial stage was
caused by electron scattering in a slightly flattened spheroidal shell.
As the shell expanded, the electron scattering optical depth
decreased, hence the intrinsic polarization declined.  A similar
interpretation could be applied to V838 Mon.  The electron scattering
wings around H$\alpha$ were sizable on February 5, but had clearly
weakened by February 9 and disappeared by February 14.  Coupled with our
discovery of an intrinsic polarization component present on February 8
but gone by February 13, this picture of an expanding, flattened
spheroidal shell could provide a viable explanation of the intrinsic
polarization observed during the 2002 outburst.

Finally, we consider the implications of these observations for future
studies of this object.  \citet{mu202} and \citet{kim2b}
 discuss different classifications
of V838 Mon, including a nova outburst, a post-AGB star, a M31-Red type
variable, and a V4332 Sgr type variable: both suggest that V838 Mon is most
similar to a V4332 Sgr type variable.  As described above, we suggest that the
geometry of the outburst, as probed by polarimetry, might be similar to that
of a nova outburst.  This suggests that the geometry of V4332 Sgr's, V838
Mon's, and nova outbursts might be similar.  It would be worthwhile to
measure the polarization of V4332 Sgr today, to verify that like V838 Mon, it
has no intrinsic polarization at a time long after outburst.  Furthermore, we
suggest that polarimetric observations immediately following the outbursts of
all future V4332 Sgr type variables be made.  Such observations would provide
an ideal testbed to correlate the geometry of each outburst, and hence help to
identify the true nature of these unique objects.

\acknowledgments

We would like to thank Dr. Kenneth H. Nordsieck for providing access to
the HPOL spectropolarimeter.  We also thank Brian Babler for his help with
various aspects of HPOL data reduction and management.  We thank the
anonymous referee for helping to improve this paper.  Support for
 observational research at
Ritter Observatory has been provided by the University of Toledo, with
technical support provided by R.J. Burmeister.  K.S.B. is a Cottrell
Scholar of the Research Corporation, and gratefully acknowledges their
support.  This research has made use of the SIMBAD database operated at
CDS, Strasbourg, France, and the NASA ADS system.

\clearpage

\begin{figure}
\epsscale{0.8}
\plotone{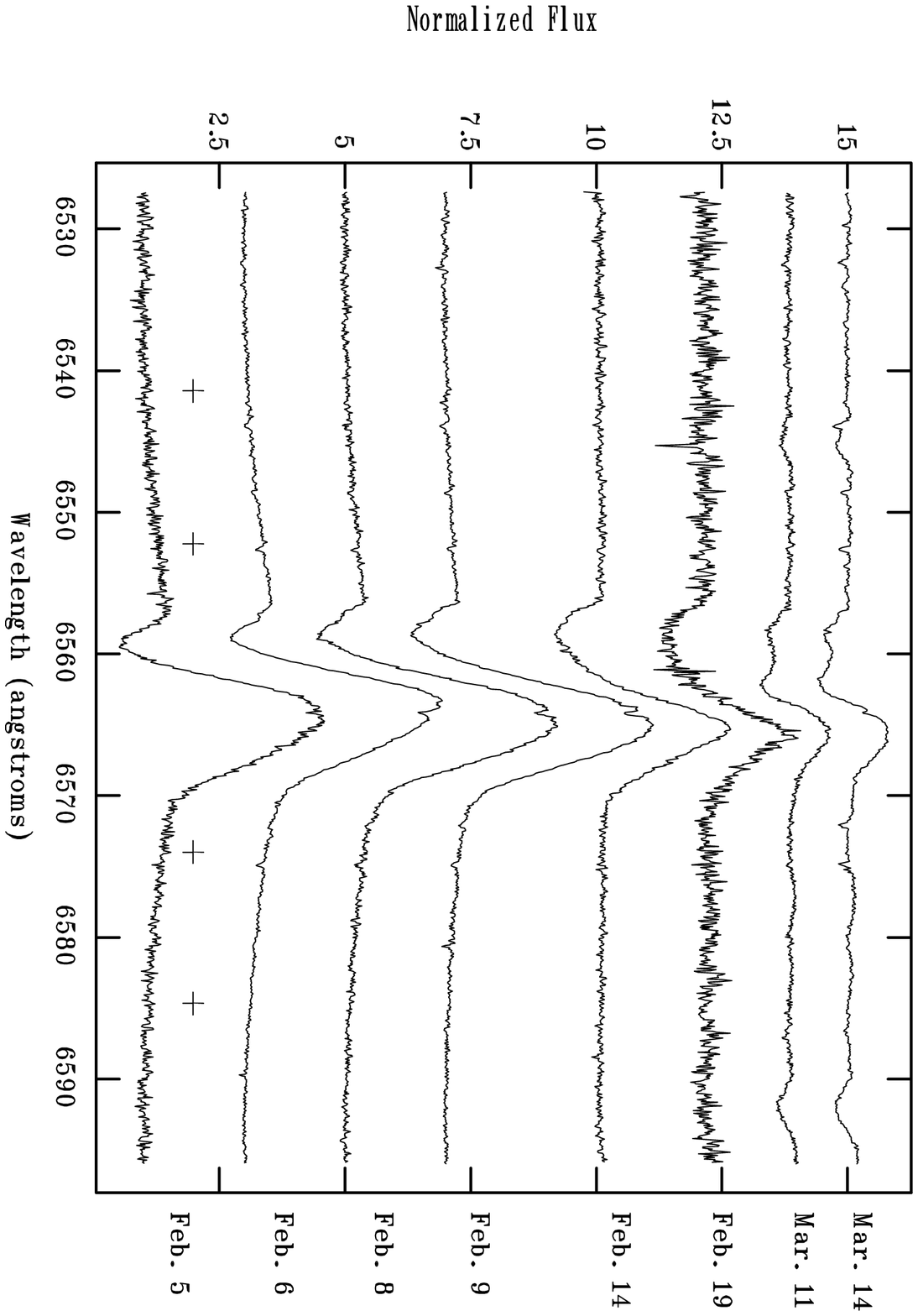}
\figcaption[f1.eps]{H$\alpha$ line profiles sorted chronologically.  From 
shorter to longer wavelengths, the tick marks denote -1000 km s$^{-1}$, 
-500 km s$^{-1}$, 500 km s$^{-1}$, and 1000 km s$^{-1}$.}
\end{figure}

\newpage
\begin{figure}
\epsscale{0.8}
\plotone{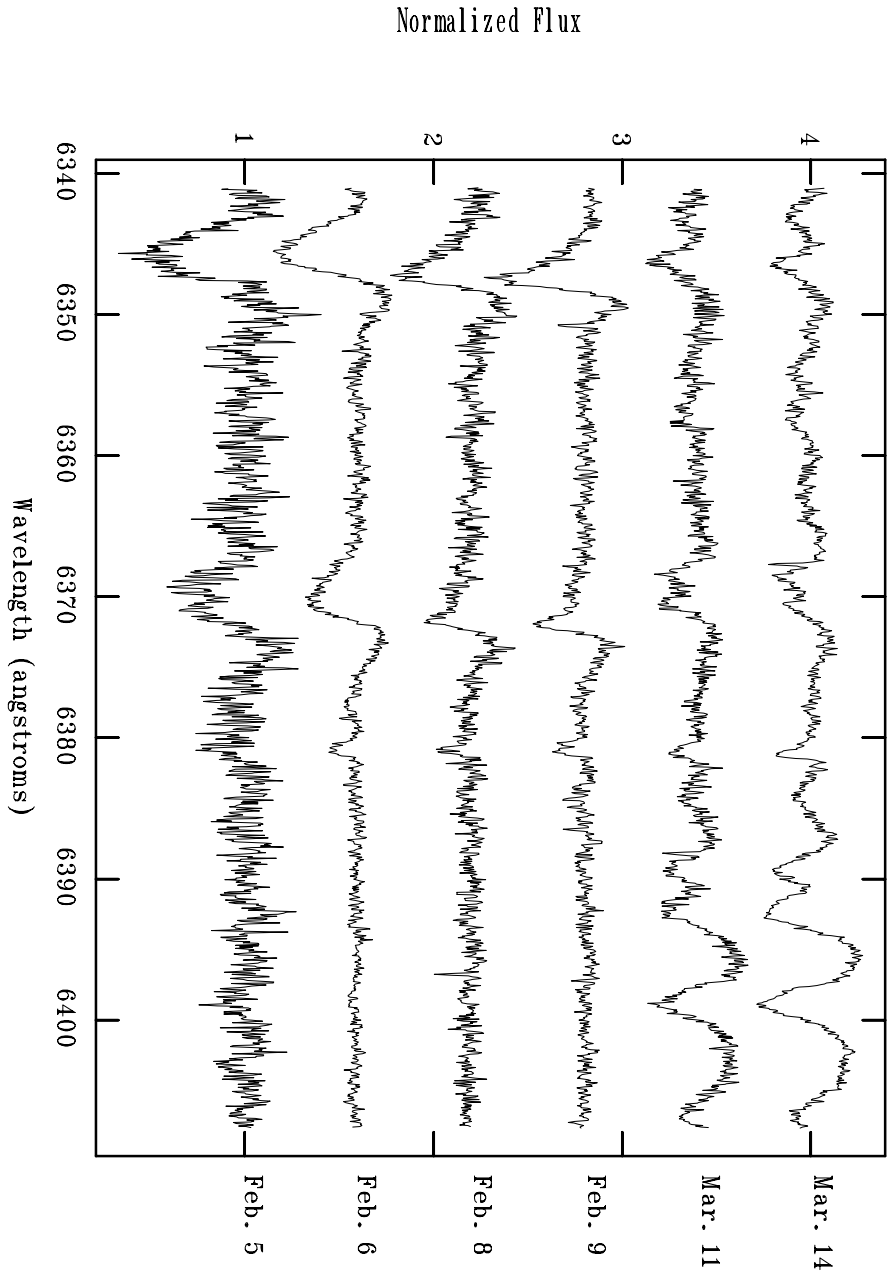}
\figcaption[f2.eps]{SiII 6347.1\AA\ and 6371.4\AA\ line profiles sorted
chronologically.}
\end{figure}

\newpage
\begin{figure}
\epsscale{0.8}
\plotone{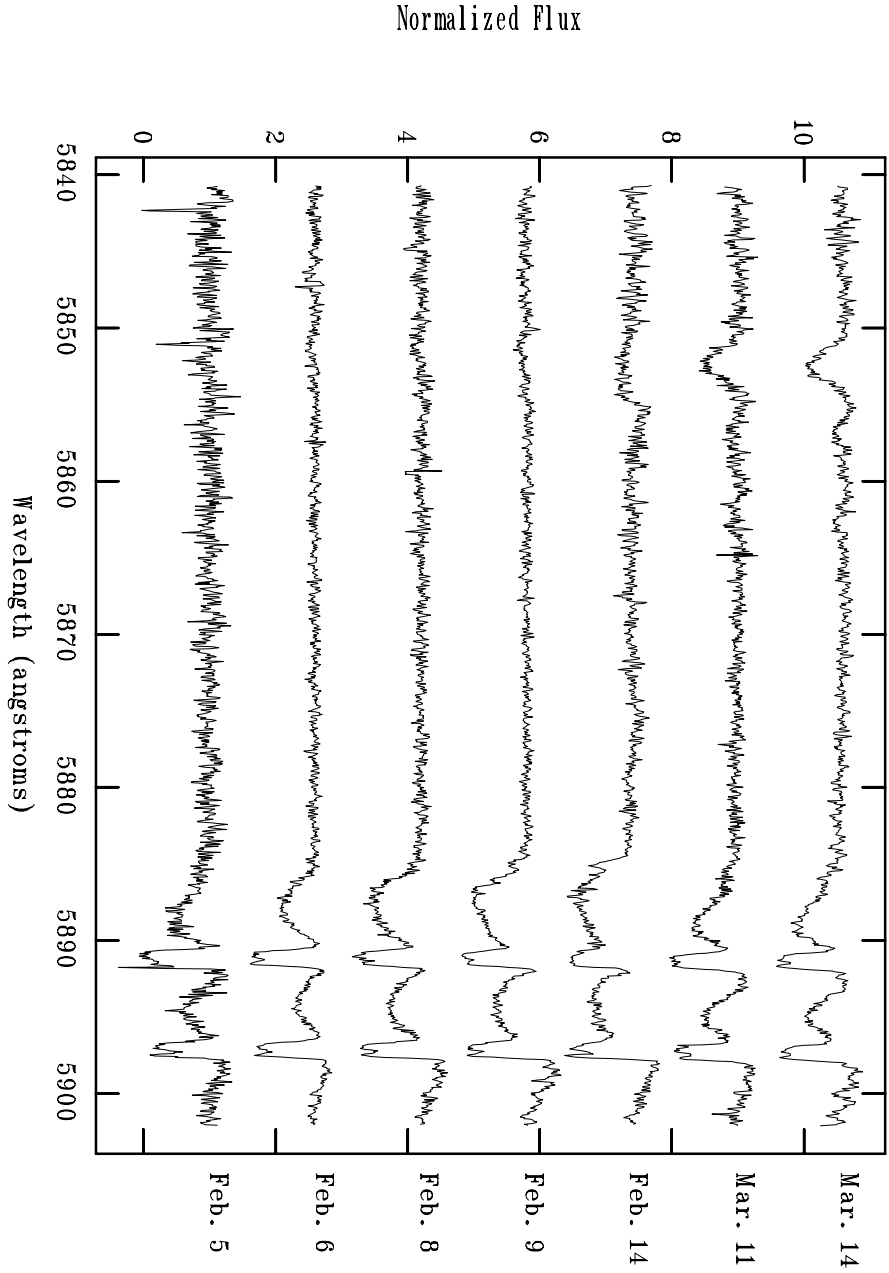}
\figcaption[f3.eps]{NaI 5889.95\AA\ and 5895.92\AA\ line profiles sorted
chronologically.  Note that the narrow interstellar line
components are superimposed on the intrinsic components.  As discussed in
section 3.1, the saturation of the interstellar components prevents the
isolation of the intrinsic components.}
\end{figure}

\newpage
\begin{figure}
\epsscale{0.7}
\plotone{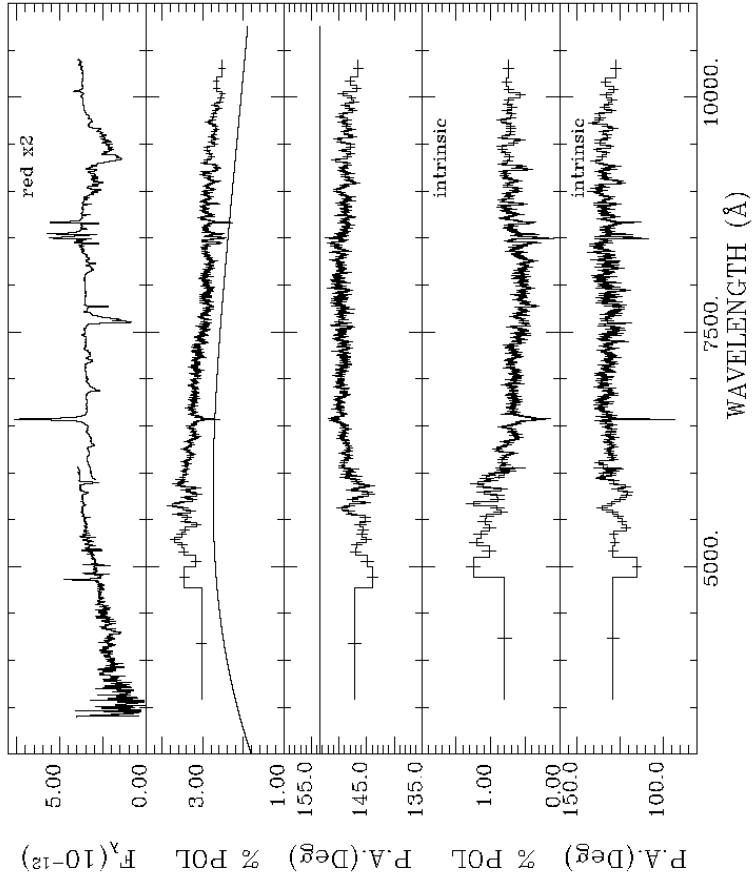}
\figcaption[f4.eps]{HPOL spectropolarimetry from February 8.  The
upper panel shows the flux, in units of ergs cm$^{-2}$ s$^{-1}$
$\AA^{-1}$ , with the red data magnified by a factor of
2.  The next two lower panels display the total polarization and
position angle, where the red data, e.g. 6000 -10500\AA, are binned to a
constant error of $0.075\%$ and blue data, e.g. 3200 -6000\AA, are
binned to a constant error of $0.12\%$.  Overplotted is the derived Serkowski
interstellar polarization component, whose parameters are given by:
$P_{max} = 2.746 \pm 0.011 \%$, $\lambda_{max} = 5790 \pm 37\AA$,
$PA = 153.43 \pm 0.12^{\circ}$, $\delta PA = 0$, and $K = 0.971$.
  The bottom two panels show the
intrinsic polarization and position angle, binned to constant errors of
$0.07\%$ and $0.10\%$ for the red and blue data respectively.  The wavelength
dependence of the intrinsic polarization is not representative of pure
electron scattering; rather, it implies the presence of an opacity source
such as hydrogen.}
\end{figure}

\newpage
\begin{figure}
\epsscale{0.8}
\plotone{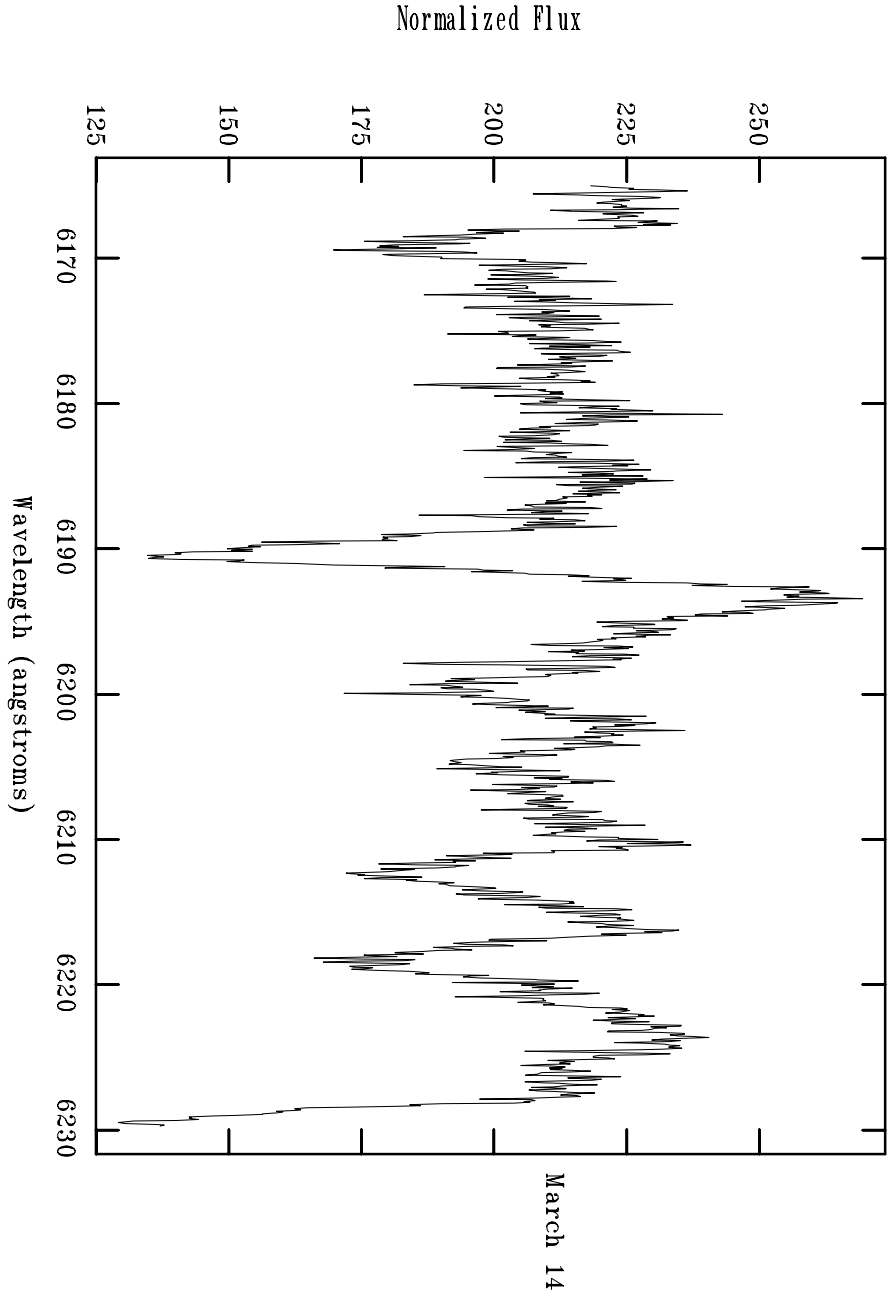}
\figcaption[f5.eps]{A strong P-Cygni profile, attributed to FeI
6191.6\AA, is shown in a non-continuum normalized spectrum from 
March 14.}
\end{figure}

\newpage
\begin{figure}
\epsscale{0.8}
\plotone{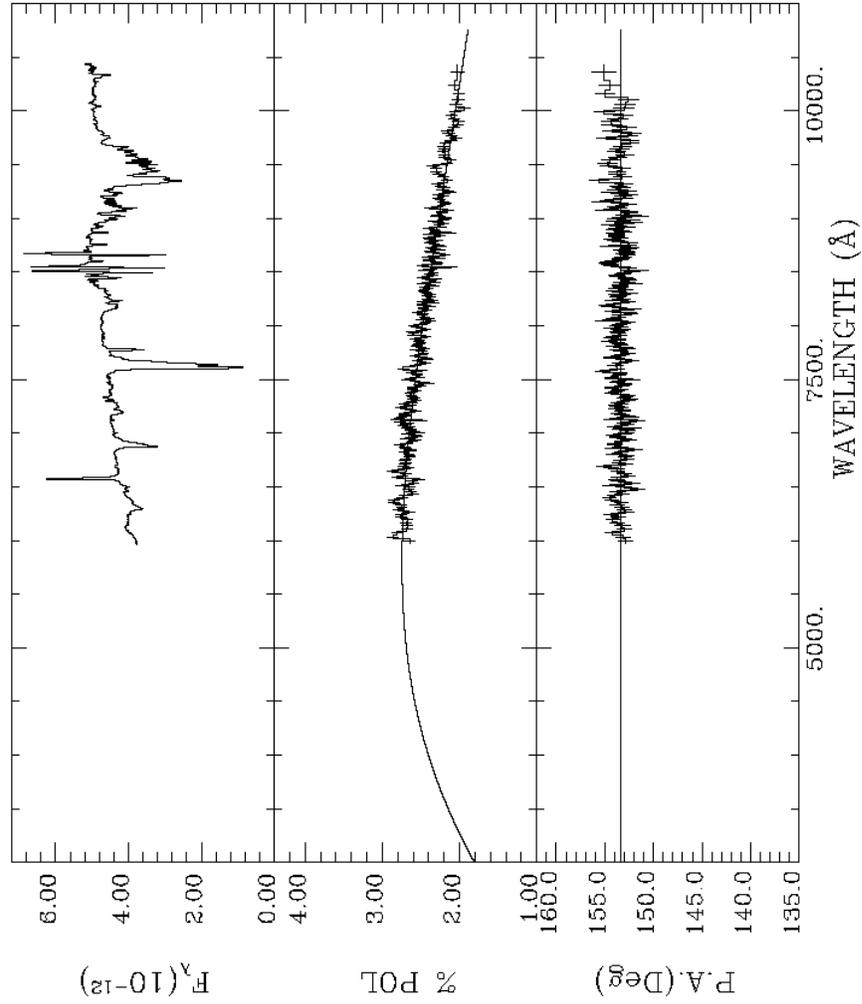}
\figcaption[f6.eps]{HPOL spectropolarimetry from February 13.
The upper panel shows the flux, in units of ergs cm$^{-2}$ s$^{-1}$$\AA^{-1}$.  The lower two panels give the
total polarization and position angle, binned to a constant error of
$0.074\%$.  The fitted interstellar polarization component is given by
the solid line.  The intrinsic polarization which was present on 8
February has clearly disappeared by February 13.}
\end{figure}

\clearpage
\normalsize
\begin{table}
\caption{}
\begin{tabular}{llccccc}

UT Date, 2002 & MJD & Observatory & SNR: H$\alpha$ & SNR: Si II & SNR: Fe I & SNR: Na I
 \\
\tableline
February 5 & 2452310.7 & Rit & 24 & 22 & \nodata & 18 \\
February 6 & 2452311.7 & Rit & 84 & 68 & \nodata & 42 \\
February 8 & 2452313.6 & Rit & 48 & 42 & \nodata & 32 \\
February 8 & 2452313.7 & HPOL & \nodata & \nodata & \nodata & \nodata \\
February 9 & 2452314.7 & Rit & 66 & 64 & \nodata & 46 \\
February 13 & 2452318.8 & HPOL & \nodata & \nodata & \nodata & \nodata \\
February 14 & 2452319.7 & Rit & 42 & \nodata & \nodata & 28 \\
February 19 & 2452324.6 & Rit & 14 & \nodata & \nodata & \nodata \\
March 11 & 2452344.6 & Rit & 60 & 64 & \nodata & 32 \\
March 14 & 2452347.6 & Rit & 104 & 84 & 78 & 34 \\

\tablecomments{Summary of observations.  Rit denotes Ritter spectroscopy
and HPOL denotes HPOL spectropolarimetry.  Multiple observations during
one night were coadded, using standard IRAF techniques, to increase the
SNR.  The Modified Julian Dates listed correspond to the midpoint of the
observations for a specific night.  The signal to noise ratios cited are the 
signal to noise ratios per resolution element, calculated in line free
regions of the spectrum.}
\end{tabular}
\end{table}
\clearpage

\clearpage
\newpage
\rotate
\begin{table}
\caption{}
\scriptsize
\begin{tabular}{lccccccccc}

Line & & & & MJD & & & & & \\

\tableline
 & 2452310.6 & 2452311.6 & 2452313.6 & 2452314.6 & 2452318.7 &
2452319.6 & 2452324.6 & 2452344.6 & 2452347.6 \\

\tableline

H$\beta$ & \nodata & \nodata & e$^{1}$ & \nodata & \nodata &
\nodata & \nodata & \nodata & \nodata \\

FeII 4923.9\AA\ & \nodata & \nodata & p$^{1}$ & \nodata & \nodata &
\nodata & \nodata & \nodata & \nodata \\

FeII 5018.4\AA\ & \nodata & \nodata & p$^{1}$ & \nodata & \nodata &
\nodata & \nodata & \nodata & \nodata \\

FeII 5169.0\AA\ & \nodata & \nodata & p$^{1}$ & \nodata & \nodata &
\nodata & \nodata & \nodata & \nodata \\

FeII 5316.2\AA\ & p & p & p$^{1}$ & p & \nodata & p? & \nodata & a & a 
\\

NaI 5889.95\AA\ & p & p & p & p & \nodata & p & \nodata & p & p \\

NaI 5895.9\AA\ & p & p & p & p & \nodata & p & \nodata & p & p \\

FeI 6191.6\AA\ & \nodata & \nodata & \nodata & \nodata &  \nodata &
\nodata & \nodata & p & p \\

SiII 6347\AA\ & p & p & p & p & \nodata &  \nodata & \nodata & p$^{2}$ &
p$^{2}$ \\

SiII 6371\AA\ & p & p & p & p & \nodata &  \nodata & \nodata & p$^{2}$ &
p$^{2}$ \\

NII 6380\AA\ & a & a & a & a & \nodata & \nodata & \nodata & a & a \\

FeI 6393.6\AA\ & \nodata & \nodata & \nodata & \nodata & \nodata & \nodata &
\nodata & p$^{2}$ & p$^{2}$ \\

H$\alpha$ & p & p & p & p & p$^{1}$ & p & p & p$^{2}$ & p$^{2}$ \\

CII 6576\AA\ & \nodata & \nodata & \nodata &  \nodata & \nodata & \nodata
& \nodata & e & e \\

CII 6583\AA\ & \nodata & \nodata & \nodata & \nodata & \nodata & \nodata
& \nodata & e & e \\

CaII 8498\AA\ & \nodata & \nodata &  p$^{1}$ & \nodata & p$^{1}$ &
\nodata & \nodata & \nodata & \nodata \\

CaII 8542\AA\ & \nodata & \nodata &  p$^{1}$ & \nodata & p$^{1}$ &
\nodata & \nodata & \nodata & \nodata \\

CaII 8662\AA\ & \nodata & \nodata &  p$^{1}$ & \nodata & p$^{1}$ &
\nodata & \nodata & \nodata & \nodata \\

Paschen & \nodata & \nodata & a$^{1}$ & \nodata & a$^{1}$ & \nodata &
\nodata &
\nodata & \nodata \\ 

HI 10049.4\AA\ & \nodata & \nodata & p?$^{1}$ & \nodata &  p?$^{1}$ &
\nodata & \nodata & \nodata & \nodata \\

\tablecomments{Observed spectral lines and their general
characteristics (p = P Cygni profile, a = absorption, e = emission).
$^{1}$ denotes line identification via low resolution HPOL
spectropolarimetry.  $^{2}$ denotes multiple absorption components
observed.}

\end{tabular}
\end{table}
\clearpage

\clearpage
\newpage
\begin{table}
\caption{}
\scriptsize
\begin{tabular}{lcccccccc}
Line & & & & MJD & & & & \\
\tableline
 & 2452310.6 & 2452311.6 & 2452313.6 & 2452314.6 & 2452319.6 &
2452324.6 & 2452344.6 & 2452347.6 \\
\tableline

H$\alpha$ (total) & -29.15 $\pm 0.75$  & -34.22 $\pm 0.25$  & -33.42
$\pm 0.43$  & -27.88 $\pm 0.26$ & -9.01 $\pm 0.15$  & -3.40 $\pm 0.26$ &
-0.66 $\pm 0.03$ & -0.73 $\pm 0.02$ \\

SiII 6347\AA\ (abs) & 1.49 $\pm 0.06$ & 1.22 $\pm 0.02$  & 0.80
$\pm0.02$ 
& 0.81 $\pm 0.02$  & \nodata & \nodata & 0.53 $\pm 0.01$
& 0.45 $\pm 0.01$ \\

SiII 6347\AA\ (em) & -0.28 $\pm 0.02$ & -0.36 $\pm 0.01$ & -0.35 $\pm
0.01$ & -0.29 $\pm 0.01$ & \nodata & \nodata & -0.11 $\pm 0.01$ & -0.12 $\pm
0.01$ \\

SiII 6371\AA\ (abs) & 0.90 $\pm 0.05$ & 0.81 $\pm 0.02$ & 0.48 $\pm
0.02$ & 0.45 $\pm 0.01$  & \nodata & \nodata & 0.43 $\pm 0.01$ & 0.37 $\pm
0.01$ \\

SiII 6371\AA\ (em) & -0.31 $\pm 0.02$  & -0.30 $\pm 0.01$  & -0.25 $\pm
0.01$ & -0.25 $\pm 0.01$  & \nodata & \nodata & -0.25 $\pm 0.01$  &
-0.23 $\pm 0.01$ \\

NII 6380\AA\ & 0.19 $\pm 0.01$  & 0.15 $\pm 0.01$  & 0.13 $\pm 0.01$  &
 0.13 $\pm 0.01$ & \nodata & \nodata & 0.12 $\pm 0.01$  & 0.13 $\pm
0.01$  \\

FeII 5316.2\AA\ (abs) & 1.98 $\pm 0.16$ & 1.51 $\pm 0.07$ & 1.65 $\pm 0.10$ &
1.42 $\pm 0.06$ & 1.03 $\pm 0.09$ & \nodata & 0.77 $\pm 0.08$ & 
0.55 $\pm 0.04$ \\

FeII 5316.2\AA\ (em) & -2.27 $\pm 0.20$ & -2.39 $\pm 0.11$ & -2.76 
 $\pm 0.17$ & -3.01 $\pm 0.10$ & -2.87 $\pm 0.20$ & \nodata &
-0.55 $\pm 0.09$ & -0.63 $\pm 0.06$ \\

\tablecomments{Equivalent width, in \AA, of selected
spectral lines for the 8 nights of Ritter observations.  Except for 
H$\alpha$ all
measurements were made on spectra smoothed with a boxcar function of
size 3.}
\end{tabular}
\end{table}
\clearpage

\end{document}